\documentclass[%
  aip,
 amsmath,amssymb,
 reprint,%
]{revtex4-1}
\usepackage{graphicx}
\usepackage{dcolumn}
\usepackage{bm}
\usepackage{color}
\usepackage{float}

\usepackage[utf8]{inputenc}
\usepackage[T1]{fontenc}
\usepackage{mathptmx}
\usepackage{subfig}
\usepackage[inline]{enumitem}
\usepackage{caption}
\newcommand{\ppx}[2]{\frac{\partial#1}{\partial#2}}

\newcommand{\ddx}[2]{\frac{d#1}{d#2}}

\renewcommand{\vec}{\mathbf}

\newcommand{\bogus}[1]{{}}

\begin{document}


\title{Bohm criterion of plasma sheaths away from asymptotic limits}

\author{Yuzhi Li}
\affiliation{Kevin T. Crofton Department of Aerospace and Ocean Engineering, Virginia Tech, Blacksburg, Virginia 24060}

\author{Bhuvana Srinivasan}
\affiliation{Kevin T. Crofton Department of Aerospace and Ocean Engineering, Virginia Tech, Blacksburg, Virginia 24060}

\author{Yanzeng Zhang}
\affiliation{Theoretical Division, Los Alamos National Laboratory, Los Alamos, New Mexico 87545}

\author{Xian-Zhu Tang}
\affiliation{Theoretical Division, Los Alamos National Laboratory, Los Alamos, New Mexico 87545}

\date{\today}

\begin{abstract}
The plasma exit flow speed at the sheath entrance is constrained by
the Bohm criterion. The so-called Bohm speed regulates the plasma
particle and power exhaust fluxes to the wall, and it is commonly
deployed as a boundary condition to exclude the sheath region in
quasi-neutral plasma modeling. Here the Bohm criterion analysis is
performed in the intermediate plasma regime away from the previously
known limiting cases of adiabatic laws and the asymptotic limit of
infinitesimal Debye length in a finite-size system, using the
transport equations of an anisotropic plasma. The resulting Bohm speed
has explicit dependence on local plasma heat flux, temperature
isotropization, and thermal force.  Comparison with kinetic
simulations demonstrates its accuracy over the plasma-sheath
transition region in which quasineutrality is weakly perturbed
and Bohm criterion applies.
\end{abstract}

\maketitle


Sheath theory has a central place in plasma physics as its original
formulation coincided with the recognition of plasma physics as a
sub-field in physics~\cite{langmuir-pra-1929,Langmuir} and it applies
to any plasma bounded by a material
boundary~\cite{lieberman-lichtenberg-book-2005,stangeby-book-2000,loarte-etal-nf-2007,hastings-garrett-book-1996,lai-book-2011}.
One of the most celebrated findings in sheath theory is the so-called
Bohm
criterion~\cite{Bohm,harrison-thompson-pps-1959,Riemann_1991,Riemann_1995,Baalrud_2011,crespo-franklin-jpp-2014,Baalrud_2015}
that predicts a threshold, the so-called Bohm speed, which would
\textcolor{black}{provide a lower bound for the plasma exit flow speed}
at the sheath entrance.  Bohm
criterion (also known as sheath criterion in the literature) is an
inequality at the sheath entrance, which can be written
as~\cite{harrison-thompson-pps-1959}
\begin{align}
  \left(\frac{\partial n_e}{\partial\phi} - Z \frac{\partial
    n_i}{\partial\phi}\right)\Big|_{\phi=\phi^{se}}\ge 0.
  \label{eq:Bohm-criterion}
\end{align}
Here $n_{e,i}$ denote the electron and ion density, respectively,
$\phi$ is the plasma potential, and the superscript $se$ labels the
sheath entrance where the plasma transitions from quasi-neutral in the
presheath to non-neutral inside the sheath. A
straightforward~\cite{Bohm,harrison-thompson-pps-1959}, but not
necessarily unique~\cite{Riemann_1991,Riemann_1995}, physics
interpretation of Bohm criterion is that Eq.~(\ref{eq:Bohm-criterion})
is required for the plasma potential to have non-oscillatory solutions
into the sheath.  This can be understood by linearizing the Poisson
equation for $\phi$ in the neighborhood of the sheath entrance where
$n_e\approx Zn_i$ remains a good approximation. The solution is of an
exponential form with the exponent imaginary if
Eq.~(\ref{eq:Bohm-criterion}) is violated, indicating an oscillatory
$\phi$ into the sheath, which would contradict the expectation of
monotonically varying $\phi$ that slows down the electrons for
ambipolarity~\cite{harrison-thompson-pps-1959}.

Traditionally, evaluation of Bohm speed from the Bohm criterion
invokes drastic simplification of plasma transport. These are
normally expressed in terms of varying $\gamma$ in the adiabatic law
$pn^{-\gamma}=constant.$ For example, $\gamma=1$ for an isothermal
plasma, $\gamma=5/3$ for an ideal plasma having three degrees of
freedom, and $\gamma=3$ for an ideal plasma constrained to one degree
of freedom.  The Bohm speed in these limiting cases then equals the
sound speed~\cite{Riemann_1995}
\begin{align}
u_{Bohm} = c_s(\gamma_e,\gamma_i)\equiv\sqrt{\left(\gamma_e T_e^{se} +
  \gamma_i T_i^{se}\right)/m_i}.\label{eq:Bohm-adiabatic}
\end{align}
\bogus{The specific choice of $(\gamma_e,\gamma_i)$ is supposed to
  give the best fit for a particular plasma under consideration.}  It
is interesting to note that although Bohm~\cite{Bohm} originally
invoked the isothermal electron approximation to realize the $\gamma_e=1$
case of Eq.~(\ref{eq:Bohm-adiabatic}), subsequent
work~\cite{godyak-etal-jap-1993} had relaxed the requirement to a
Boltzmann distribution for the electron density, $n_e = n_0
\exp\left(e\phi/ T_e^*\right),$ with $\phi$ the plasma potential and
$T_e^*$ an effective or screening temperature, the latter of which is
interpreted as what Langmuir probes are supposed to measure.
\bogus{More recently it is found~\cite{bi_max} that in the limit of a
  collisionless sheath with an upstream collisional plasma, the
  truncated bi-Maxwellian model for the sheath/presheath electrons,
  which captures the strong $T_e$ variation in the sheath/presheath,
  is able to tie the screening temperature to the electron temperature
  at the sheath entrance.}

It was recognized early on~\cite{harrison-thompson-pps-1959} that
\textcolor{black}{transport in the neighorhood of the sheath} can greatly
complicate the physics constraint set by the Bohm criterion.  A large body
of
work~\cite{Allen_JPD_1976,Bissell_1987,Riemann_1991,Riemann_1995,Riemann-PoP-2006}
has since been devoted to the development of the so-called kinetic
Bohm criterion, which is obtained by integrating the kinetic equation
for $n_{i,e}$ in Eq.~(\ref{eq:Bohm-criterion}). The standard
expression bears the form
\begin{align}
  \frac{1}{m_i}\int d^3\vec{v} \frac{f_i(\vec{v})}{v_z^2} \le - \frac{1}{m_e}\int d^3\vec{v}
  \frac{1}{v_z}\frac{\partial f_e(\vec{v})}{\partial v_z} \label{eq:kinetic-Bohm-criterion}
\end{align}
with $m_i (m_e)$ the ion (electron) mass, \textcolor{black}{$f_i
  (f_e)$ the ion (electron) distribution function}, and velocity $v_z$
which is normal to the wall in an unmagnetized plasma or parallel to
the magnetic field in a magnetized plasma. A recent
debate~\cite{Baalrud_2011,Riemann-PSST-2012,Baalrud-Hegna-PSST-2012}
highlighted a profound disconnect between (1) the conventional theory
of the Bohm criterion and (2) the practical needs in plasmas that we
normally encounter. Specifically, the Bohm criterion like in
Eq.~(\ref{eq:kinetic-Bohm-criterion}) was derived in the asymptotic
limit of $\lambda_D/L \rightarrow 0$~\cite{Riemann-PSST-2012} with
$\lambda_D$ the Debye length and $L$ the plasma size, while
plasmas of practical interest are frequently away from this asymptotic
limit~\cite{Baalrud-Hegna-PSST-2012}. The underlying challenge echoes
back to an earlier
discussion~\cite{godyak-sternberg-ieee-1990,franklin-ieee-2002,kaganovich-pop-2002,godyak-sternberg-ieee-2003,Riemann-JPD-2003,Franklin-JPD-2004}
on where the sheath entrance or edge resides, an intimately connected
issue since that is where the Bohm criterion is supposed to be
applied.

The complication is that between the quasineutral plasma and the
non-neutral Debye sheath in a plasma away from the asymptotic limit of
$\lambda_D/L\rightarrow 0,$ there is usually a transition layer in
which the quasineutrality is weakly violated, and the plasma flow and
potential (and its gradient and hence electric field) can vary
gradually~\cite{franklin_ockendon_1970,sternberg-godyak-ieee-2003,Riemann-JPD-2003,Franklin-JPD-2004}.
Matched asymptotic analysis of a simplified plasma model with
isothermal electrons and cold ions, reveals that the plasma ion flow
actually crosses the classically defined Bohm speed~\cite{Bohm}
$u_{Bohm}=\sqrt{T_e/m_i}$ somewhere inside this transition
layer~\cite{franklin_ockendon_1970,Franklin-JPD-2004}.  This is
consistent with the straightforward interpretation of the Bohm
criterion as given in Eq.~(\ref{eq:Bohm-criterion}) by Harrison and
Thompson~\cite{harrison-thompson-pps-1959} that (1) it offers no
meaningful constraint in the quasineutral region because $n_e\approx
Zn_i$ and Poisson's equation is not used for evaluating $\phi;$ (2) it
does not apply in the Debye sheath in the sense of Langmuir and
Tonks~\cite{langmuir-pra-1929,Langmuir} since $n_e$ grossly differs
from $Zn_i,$ and (3) it does impose a constraint, as we shall show in
this Letter, on the ion flow speed over the {\em
  spatially extended transition region}, as opposed to a {\em sharp
  transition boundary}, over which quasineutrality is mildly perturbed
so charge density gradient is the dominant term upon linearization of
Poisson's equation. This last point implies a Bohm speed that should
{\em vary inside this transition region}.

In this Letter, we derive an expression for the Bohm speed away from
the previously known asymptotic limits, that elucidates the distinct
roles of various transport physics, including heat flux, collisional
isotropization, and thermal force for both electron and ion transport.
Its explicit dependence on plasma transport and local electric field
suggests a spatially varying Bohm speed over a transition region in
which quasineutrality is weakly perturbed. This is confirmed by
first-principle kinetic simulations over a range of plasma
collisionality. To our knowledge, this is the first time that {\em a
predictive formula for Bohm speed has been shown to be quantitatively
accurate in the intermediate plasma regime that is away from the
 limiting cases of adiabatic laws and the asymptotic limit of
$\lambda_D/L\rightarrow 0.$}


The nature of plasma transport in the sheath/presheath region is
governed by the sheath Knudsen number $K_n,$ which is the ratio
between plasma mean-free-path $\lambda_{mfp}$ and the Debye length
$\lambda_D.$ In cases of most interest, $K_n >1$ or $K_n\gg 1.$
 A consequence is that within the Knudsen layer, which is defined as one
mean-free-path ($\lambda_{mfp}$) within the wall, streaming loss and
the associated decompressional cooling would induce robust temperature
anisotropy~\cite{Tang_2011}, $T_\parallel < T_\perp.$ The parallel
degree of freedom is along the magnetic field, or in an unmagnetized
plasma the plasma flow direction, which is normal to the wall surface.
{Due to the anisotropic nature of the plasma,
the mean-free-path is defined as $\lambda_{mfp} \equiv v_{the}/\nu_{ei}$
with $v_{the} = \sqrt{T_{e\parallel}/m_i}$ the electron thermal velocity
and $\nu_{ei}$ the electron-ion collision frequency in an anisotropic plamsa
given by Eq.~(\ref{eq:collision-frequency}).}
Here we will focus on a magnetized plasma, with a
uniform magnetic field normal to the wall ($T_\parallel=T_x$) and
$y$ signifying a perpendicular direction ($T_\perp = T_y$).  The
plasma transport equations that directly enter the Bohm speed
evaluation include the species continuity equation, momentum equation,
and energy equation, all in the parallel or $x$ direction, which in
the neighborhood of the sheath entrance, take the form,
\begin{subequations}
  \begin{equation}
    \ppx{n_e u_{ex}}{x} = 0; \,\,\,
    \ppx{n_i u_{ix}}{x} = 0, \label{eq:transport-equation-continuity}
  \end{equation}
  \begin{equation}
    \ppx{n_eT_{ex}}{x} = en_e\ppx{\phi}{x}-\alpha n_e\ddx{T_{ex}}{x}, \label{eq:transport-equation-e-momentum}
  \end{equation}
\begin{equation}
    n_im_iu_{ix}\ppx{u_{ix}}{x}+\ppx{n_iT_{ix}}{x} = -Zen_i\ppx{\phi}{x}+\alpha n_e\ddx{T_{ex}}{x}, \label{eq:transport-equation-i-momentum}
  \end{equation}
\begin{equation}     
  	n_eu_{ex}\ppx{T_{ex}}{x}+2n_eT_{ex}\ppx{u_{ex}}{x}+\ppx{q_n^e}{x}  = Q_{ee}+Q_{ei}, \label{eq:transport-equation-Te}
\end{equation}
\begin{equation}
    n_iu_{ix}\ppx{T_{ix}}{x}+2n_iT_{ix}\ppx{u_{ix}}{x}+\ppx{q_n^i}{x}= Q_{ii}. \label{eq:transport-equation-Ti}
\end{equation}
\label{eq:transport-equation}
\end{subequations}
Here we have ignored the electron inertia and a net plasma current into the wall,
$\alpha$ is the thermal force coefficient, $q_n^{e,i}$ are the heat flux of $x$-degree of freedom in the $x$ direction,
\begin{align}
q_n \equiv \int m \left(v_x - u_x\right)^3fd^3\vec{v},
\end{align}
and ($Q_{ee}, Q_{ei}, Q_{ii}$) are temperature isotropization terms in
an anisotropic plasma, which in high collisionality
limit~\cite{Chodura_1971} have the form,
\begin{align}
  Q_{ee} \approx \frac{\sqrt{2}}{2} Q_{ei}= & 8n_e \nu_{ee}T_{ey}\frac{T_{ex}}{T_{ey}-T_{ex}}\Big[-3+\Big(3\sqrt{\frac{T_{ex}}{T_{ey}-T_{ex}}}+  \nonumber \\
    & \sqrt{\frac{T_{ey}-T_{ex}}{T_{ex}}}\Big)\arctan{\sqrt{\frac{T_{ey}-T_{ex}}{T_{ex}}}}\Big],
\end{align}
with the collision rate
\begin{align}
  \label{eq:collision-frequency}
\nu_{ee} = \frac{n_e}{n_i}\frac{\nu_{ei}}{\sqrt{2}}
=\frac{\sqrt{\pi}}{2}n_e \frac{e^4}{(4 \pi \epsilon_0)^2}
\frac{\ln{\Lambda}}{\sqrt{m_eT_{ex}}T_{ey}}.
\end{align}

The evaluation of the Bohm speed can now be performed following
Ref.~\cite{tang2016}. Combining the electron continuity equation,
momentum equation, and energy equation, we can substitute out the
$\partial T_{ex}/\partial x$ and $\partial u_{ex}/\partial x$ terms
and find that in the neighborhood of the sheath entrance where $\phi$
is a monotonically varying function of $x,$
\begin{align}
  \ppx{n_e}{\phi} =
  \frac{en_e}{(3+2\alpha)T_{ex}}+\frac{1+\alpha}{(3+2\alpha)u_{ex}T_{ex}}\Big(\ppx{q_n^e}{\phi}+\frac{Q_{ee}+Q_{ei}}{E}\Big)
  \label{eqn:dnedphi}
\end{align}
where $E=-\partial\phi/\partial x$ is the electric field.
In contrast, the ion inertia must be retained in a similar analysis of the ion continuity, momentum, and energy
equations, and the result is
\begin{align}
 \ppx{n_i}{\phi} = \frac{1}{3u_{ix}T_{ix}-m_iu_{ix}^3}\left(\ppx{q^i_n}{\phi}+
 \frac{Q_{ii}}{E}\right)-\frac{Zen_i-\alpha n_e{\partial T_{ex}}/{\partial
     \phi}}{3T_{ix}-m_iu_{ix}^2}.
  \label{eqn:dndphi}
\end{align}
Substituting Eqs.~(\ref{eqn:dnedphi}) and (\ref{eqn:dndphi}) into
Eq.~(\ref{eq:Bohm-criterion}), and rearranging terms, we find that the
Bohm criterion provides a lower bound for the plasma exit flow speed,
\begin{align}
u_{ix}^{se} \ge u_{Bohm}
\end{align}
with
\begin{align}
  u_{Bohm} \equiv
  \sqrt{\frac{Z \beta T_{ex}^{se} + 3 T_{ix}^{se}}{m_i}}, \label{eq:Bohm-speed-def}
\end{align}
and
\begin{align}
\quad \beta \equiv
\cfrac{3-\cfrac{3+2\alpha}{Ze\Gamma_i^{se}}\left(\cfrac{\partial
    q_n^i}{\partial \phi} +
  \cfrac{Q_{ii}}{E}\right)+\cfrac{\alpha}{e\Gamma_e^{se}}\left(\cfrac{\partial
    q_n^e}{\partial
    \phi}+\cfrac{Q_{ee}+Q_{ei}}{E}\right)}{1+\cfrac{1+\alpha}{e\Gamma_e^{se}}\left(\cfrac{\partial
    q_n^e}{\partial \phi}+\cfrac{Q_{ee}+Q_{ei}}{E}\right)}. \label{eq:gamma-def}
\end{align}
Here $\Gamma_{e,i}=n_{e,i}u_{ex,ix},$
and all quantities on the right hand side of
Eq.(\ref{eq:gamma-def}) are evaluated locally at the sheath entrance, which is interpreted
here as the plasma-to-sheath transition region where quasineutrality is weakly violated.

The Bohm speed defined in
Eqs.~(\ref{eq:Bohm-speed-def},\ref{eq:gamma-def}) takes into account
the known collisional transport physics. It recovers the collisionless
sheath/presheath limit previously
found in Ref.~\cite{tang2016}, which is obtained by setting $\alpha,
Q_{ee},Q_{ei},$ and $Q_{ii}$ to zero,
\begin{align}
\beta = \left(3 - \cfrac{3}{Z e\Gamma_i^{se}} \cfrac{\partial
    q_n^i}{\partial\phi}\right)\Biggm/\left(1 + \cfrac{1}{e\Gamma_e^{se}}\cfrac{\partial
    q_n^e}{\partial\phi}\right)
\end{align}
A particularly interesting limit is $L \gg \lambda_{mfp} \gg
\lambda_D$ so the upstream plasma is a Maxwellian. The
presheath/sheath electrons follow a truncated bi-Maxwellian due to the
trapping effect of the ambipolar electrostatic potential, which gives
rise to an electron heat flux that satisfies $\partial
q_n^e/\partial\phi = 2e\Gamma_e^{se}.$~\cite{bi_max} Ignoring the much
smaller ion heat flux, one then finds $u_{Bohm}=\sqrt{\left(T_{ex} + 3
  T_{ix}\right)/m_i}$ because of the dominant contribution from the
electron heat flux term.~\cite{tang2016} This strikes a remarkable
but superficial coincidence with the Bohm speed expression in
Eq.~(\ref{eq:Bohm-adiabatic}) for $c_s(\gamma_e=1, \gamma_i=3).$

The full expression in Eq.~(\ref{eq:gamma-def}) allows us to quantify
the transport physics effect on Bohm speed over a wide range of plasma
collisionality. Perhaps the subtlest factor is the collisional
temperature isotropization. Naively, one would expect $Q_{ee}$ to be
small when plasma collisionality is either strong in which case $T_y-
T_x$ vanishes, or weak in which case $\nu_{ee}$ becomes negligibly
small.  This can be quantitatively assessed by expanding $Q_{ee}$ in
the small parameter of $X\equiv \sqrt{(T_{ey}-T_{ex})/T_{ex}}.$ To
leading order in $X,$ the collisional closure of
  Chodura and Pohl~\cite{Chodura_1971} predicts
\begin{align}
Q_{ee} = \frac{32}{15} n_e \nu_{ee}T_{ey} X^2.
\end{align}
The collisional temperature isotropization enters the Bohm speed with
normalization by the electron flux and electric field at the
sheath entrance,
\begin{align}
  \frac{Q_{ee}+Q_{ei}}{e\Gamma_e E} & \approx (1+\sqrt{2})\frac{32}{15} \frac{n_e\nu_{ee} T_{ey} X^2}{e n_e u_{ex} E} \nonumber \\
  & = (1+\sqrt{2})\frac{32}{15} \frac{v_{th,e}}{u_{ex}^{se}} \frac{\lambda_D}{\lambda_{mfp}} \frac{T_{ey}}{e E \lambda_D} X^2 \nonumber \\
  & \approx (1+\sqrt{2})\frac{32}{15} \sqrt{\frac{T_{ex}^{se}}{\beta T_{ex}^{se} + 3 T_{ix}^{se}}}
  \frac{\sqrt{m_i/m_e}}{K_n} \frac{T_{ey}}{\lambda_D e E} X^2. \label{eq:Qee-E-expansion}
\end{align}
In the collisionless sheath limit $K_n\rightarrow\infty$ but all the
other terms are bounded so
\begin{align}
  \lim_{K_n \rightarrow \infty}\frac{Q_{ee}+Q_{ei}}{e\Gamma_e E} = 0,
\end{align}
which is the limiting result to be expected.  In the intermediate
regime of finite collisionality, different offsetting physics can
produce an order-unity $(Q_{ee}+Q_{ei})/e\Gamma_e E$ that has an indispensable
role in setting the Bohm speed.  In the high collisionality regime,
which is denoted by $K_n < \sqrt{m_i/m_e},$ the small but still finite
temperature anisotropy is the offsetting factor that produces a
$(Q_{ee}+Q_{ei})/e\Gamma_e E \sim O(1).$ With a decreasing collisionality so
$K_n>\sqrt{m_i/m_e}$ but not too much greater, there are two
offsetting factors coming into play. The first is the familiar
temperature anisotropy, which can be enhanced by an order of
magnitude. The second is a much reduced electric field at the sheath
entrance, which can boost the factor $T_{ey}/\lambda_D e E.$
Overall, one finds that for a range in which $K_n>\sqrt{m_i/m_e},$
the two effects can offset a large $K_n,$ so $(Q_{ee}+Q_{ei})/e\Gamma_e E$ remains order unity
and hence has an important role in setting the Bohm speed.

Next we deploy first-principle kinetic simulations to verify the Bohm
speed of Eq.~(\ref{eq:Bohm-speed-def},\ref{eq:gamma-def}) and quantify
the relative importance of various transport physics under
consideration.  The VPIC~\cite{VPIC} simulations are for a slab plasma bounded by
absorbing walls at $x=0$ and $x=L.$ The loss at the wall is balanced
by a plasma source in the middle $x\in [3/8L, 5/8L],$ as a way to
mimic the upstream source for the scrape-off layer plasma in a
tokamak. Other specifics include $L=256\lambda_D, N=10000$ markers per
cell, $Z=1, m_i/m_e=1836$. The source temperatures $T_{e0}=T_{i0}$ and
the background or initial plasma density $n_0$ will be varied so the
sheath Knudsen number $K_n\in [20,5000].$ The uniform
  magnetic field is strong so the plasma beta is much less than
  unity, $\sim 1$\%. At the sheath entrance
$K_n^{se}$ would be smaller, but proportional to $K_n.$ There are
three essential points we will focus on
here.~\cite{VPIC-simulation-comment}

\begin{figure}[h!]
	\includegraphics[width=0.5\textwidth]{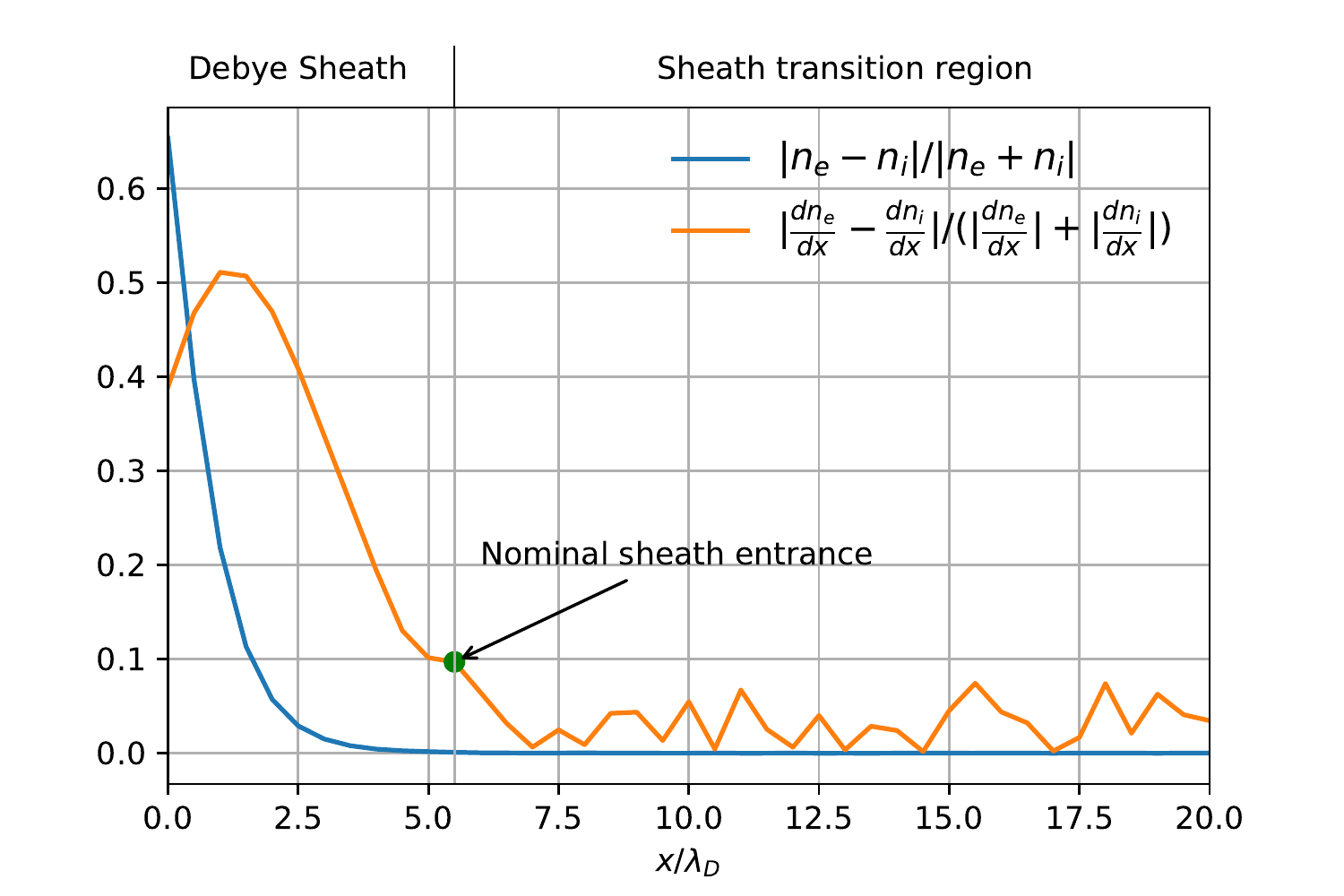}
	\caption{The normalized net charge density $\overline{\rho}$,
          and the fractional charge density gradient
          $\overline{\partial{\rho}/\partial{x}}$ for $K_n = 200.$}
	\label{fig:sheath-transition}
\end{figure} 

The first point is on the sheath entrance, which for a plasma away
from the asymptotic limit of $\lambda_D/L\rightarrow 0,$ covers a
transition region in which deviation from quasineutrality is small but
finite.  The transition into the sheath can be most obviously assessed
by fractional charge density $|n_e-n_i|/(n_e+n_i),$ but a more
sensitive measure for Bohm criterion is $|\partial n_e/\partial x -
\partial n_i/\partial x|/\left(|\partial n_e/\partial x| + |\partial
n_i/\partial x|\right).$ In Fig.~\ref{fig:sheath-transition}, one can
see that with the PIC noise of $N=10000$ markers per cell, we can
reliably position the edge of the sheath transition region to $x >
5\lambda_D$ using the charge density gradient, while the charge
density itself gives a sensitivity to $x=2.5\lambda_D.$ The simulation
data of the fractional charge density gradient, which has higher
sensitivity, is consistent with a sheath transition region over which
the violation of quasineutrality is small, and proceeds gradually
towards the non-neutral Debye sheath.

\begin{figure}[h!]
  \includegraphics[width=0.5\textwidth]{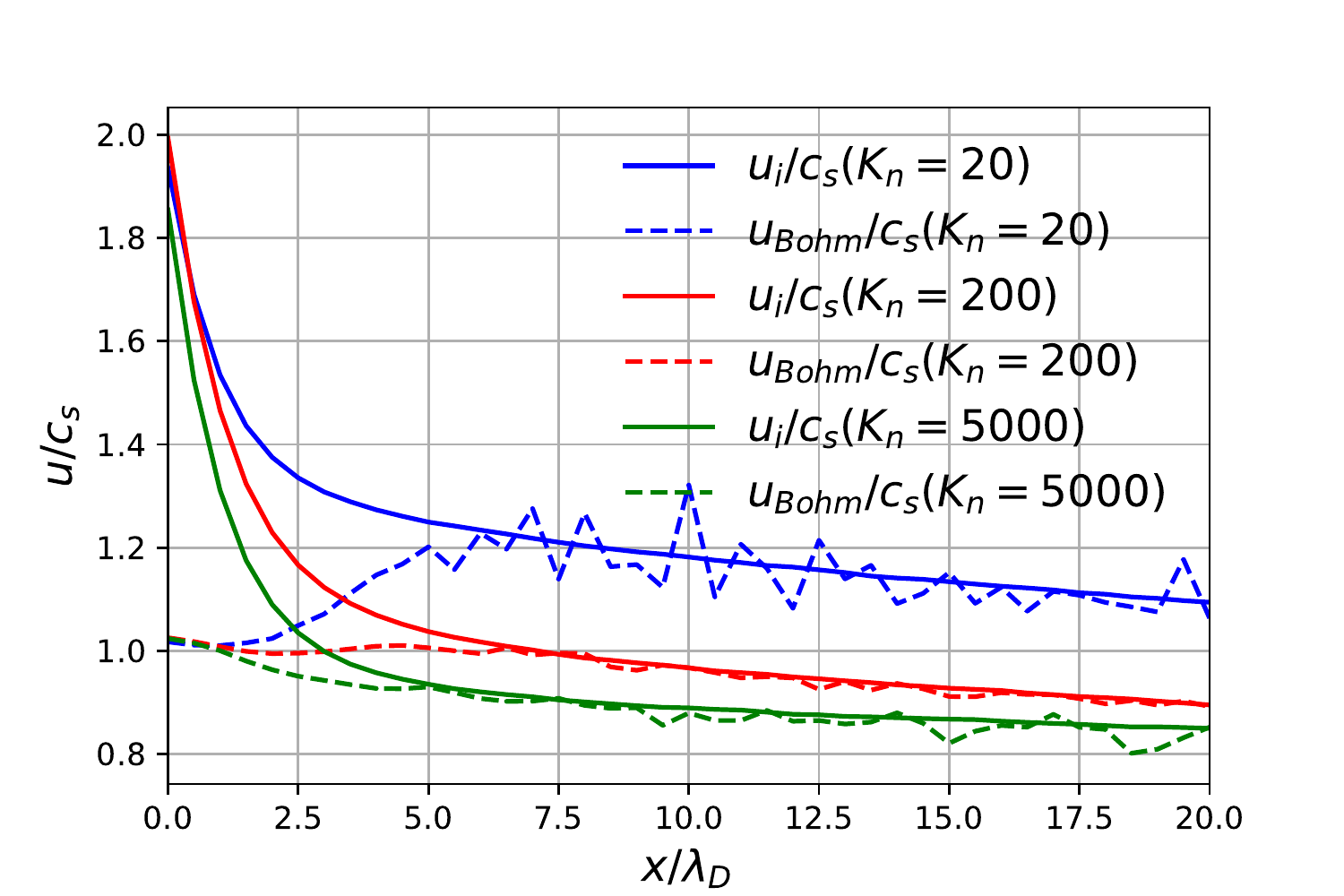} 
    \caption{Ion exit flow speed from simulation data and Bohm speed
      calculated from Eq.~(\ref{eq:Bohm-speed-def},\ref{eq:gamma-def})
      normalized by $c_s(\gamma_e=1,\gamma_i=3)$ in
      Eq.~(\ref{eq:Bohm-adiabatic}) over distance from wall for
      $K_n=20, 200, 5000.$ The breakdown of $u_{Bohm}$ from
      Eq.~(\ref{eq:Bohm-speed-def},\ref{eq:gamma-def}) for Bohm speed
      is an accurate indication of transitioning into non-neutral
      Debye sheath.}
    \label{fig:Bohm-speed-transition}
\end{figure}

The second point is that over the {\em spatially extended sheath
  transition region}, the Bohm criterion should be applicable with a
high degree of accuracy that is measured by the fractional change in
charge density gradient.  In Fig.~\ref{fig:Bohm-speed-transition}, we
contrast the ion flow speed from VPIC simulations, with the Bohm speed
from Eq.~(\ref{eq:Bohm-speed-def},\ref{eq:gamma-def}), as a function
of position from the wall. Here, in evaluating the Bohm speed, we
compute all individual terms in Eq.~(\ref{eq:gamma-def}) using the
VPIC simulation data. Since the terms in Eq.~(\ref{eq:gamma-def})
involve higher-order velocity moments and their derivatives, we deploy
time-averaging (but not spatio-averaging) over a long period in which
the plasma has reached steady-state. This overcomes the constraint of
the normal PIC noise level of $1/\sqrt{N}$ with $N$ the particle
markers per cell. The inherent PIC noise has been sufficiently
suppressed that we can see a clear sheath transition region over which
the ion flow speed closely follows the Bohm speed of
Eq.~(\ref{eq:Bohm-speed-def},\ref{eq:gamma-def}) in
Fig.~\ref{fig:Bohm-speed-transition}.  Further into the Debye sheath,
the ion flow speed diverges from the locally evaluated Bohm speed to
become significantly greater, as expected.
Further away from the Debye sheath and wall, the theoretical expectation is that
Eq.~(\ref{eq:Bohm-speed-def},\ref{eq:gamma-def}) would set a local
Bohm speed as long as it is still within the transition region where
quasineutrality is weakly perturbed.  It must be emphasized that in
the quasineutral region, the Bohm criterion as of
Eq.~(\ref{eq:Bohm-criterion}) is not a viable concept, so
Eq.~(\ref{eq:Bohm-criterion}) no longer produces a physically
meaningful speed to constrain the ion flow.

\begin{table} 
  \caption{\label{table-Bohm} Sheath quantities (columns) around the
    nominal sheath entrance ($x/\lambda_D$) for nominal~\cite{nominal-def-comment}
    $K_n=(20, 200, 5000)$ cases (rows), with $K_n^{se}$ the local Knudsen
    number.}
\begin{ruledtabular}
\begin{tabular}{cccccccccc}
 $K_n^{se}$ & $\cfrac{x}{\lambda_D}$ & $\cfrac{u_{ix}}{c_s}$  &  $\cfrac{u_{Bohm}}{c_s}$ & $\cfrac{1}{e\Gamma_e^{se}} \cfrac{\partial q_n^{e}}{\partial\phi}$ & $\cfrac{Q_{ee}^{se}+Q_{ei}^{se}}{e\Gamma_{e}^{se}E^{se}}$ & $\cfrac{1}{e\Gamma_i^{se}} \cfrac{\partial q_n^{i}}{\partial\phi}$ & $\cfrac{Q_{ii}^{se}}{e\Gamma_{i}^{se}E^{se}}$ & $\alpha^{se}$\\
 \hline
  5.26 & 5.0  & 1.25 & 1.20 & -1.63 & 1.48 & -0.05 &  0.18  & 0.59 \\
  50.0 & 5.5  & 1.02 & 1.00 & -2.20 & 3.55 & -0.22 &  0.38  & 0.45 \\
  1932 & 5.0  & 0.94 & 0.93 &  0.23 & 1.17 &  0.52 &  0.07  & 0.04     
\end{tabular}
\end{ruledtabular}
\end{table}

The third point is on the relative importance of various transport
physics in setting the Bohm speed. The transport under examination is
collisional by nature, and includes thermal force, heat flux, and
collisional temperature isotropization, for both electrons and
ions. We are particularly interested in how these dependencies vary
(a) with collisionality $K_n$ and (b) over space in the transition
layer of a given $K_n.$ For (a), we contrast the ion flow speed with
the Bohm speed at a nominal sheath entrance point for different
nominal $K_n$ cases. The terms in Eq.~(\ref{eq:gamma-def}) are
computed from simulation data and separately tabulated in
Table~\ref{table-Bohm} to quantify their relative
importance~\cite{closure-comment}. Also shown are the local sheath
Knudsen number $K_n^{se},$ ion exit flow speed $u_{ix}$ directly from
simulations, and $u_{Bohm}$ computed from
Eq.~(\ref{eq:Bohm-speed-def},\ref{eq:gamma-def}) using the tabulated
data for each case. Both $u_{ix}$ and $u_{Bohm}$ are normalized by
$c_s(\gamma_e=1,\gamma_i=3)$ from Eq.~(\ref{eq:Bohm-adiabatic}), using
$T_{ex}^{se}$ amd $T_{ix}^{se}$ from the simulations. The electron
thermal flux enters through a divergence in the energy equation, so it
is a dominant term in sheath analysis~\cite{tang2016}. This is clearly
indicated by the data, with additional subtleties in the high $K_n$
limit that the whistler instability driven by trapped
electrons~\cite{guo-tang-prl-2012b} can modify the parallel electron
thermal conduction flux in a magnetized plasma, and magnetic field
strength modulation on sheath scale can also modify the parallel
thermal flux~\cite{guo-tang-prl-2012a}.  As previously discussed after
Eq.~(\ref{eq:Qee-E-expansion}), the collisional electron temperature
isotropization has an equally important role that is further aided by
the decreasing local electric field (in magnitude) as $K_n$
increases. An accurate $E^{se}$ was previously found by
Kaganovich~\cite{kaganovich-pop-2002} to be important for matching the
sheath solution to the quasineutral plasma in a two-scale analysis,
here we find that it enters explicitly in the Bohm speed as well.
Table~\ref{table-Bohm} also reveals that despite the mass ratio in a
hydrogen plasma, ion heat flux and ion temperature isotropization can
have a small but appreciable contribution to the Bohm speed. Finally,
the thermal force coefficient $\alpha^{se}$ is directly measured from
simulation data, and one can verify that Bohm speed has a very weak
dependence on $\alpha^{se}$ for $\alpha$ less than or equal to the
Braginskii value.  For (b), we have the remarkable finding that the
heat flux gradient and collisional temperature isotropization terms
vary substantially in the sheath transition layer, but together they
produce a Bohm speed from
Eq.~(\ref{eq:Bohm-speed-def},\ref{eq:gamma-def}) that agrees
accurately with simulated ion flow over space in
Fig.~\ref{fig:Bohm-speed-transition}. The detailed data for such a
comparison is given in the supplemental material for the $K_n=200$
case.

In conclusion, we have derived an expression for the Bohm speed that
is accurate over a broad range of plasma collisionality. The Bohm
speed is derived from the transport equations of an anisotropic
plasma, which is expected for the sheath transition problem. This
expression is verified by comparison with first-principle kinetic
simulations, within the bounds set by the PIC noise. Of particular
interest is that the Bohm speed thus formulated applies to the sheath
transition region in which the quasineutrality is weakly perturbed.
This, to our knowledge, is the first time that a predictive formula
for Bohm speed has been shown to be quantitatively accurate in the
intermediate plasma regime, which is away from the known limiting
cases and the asymptotic limit of $\lambda_D/L\rightarrow 0.$ Our
analysis can be readily extended for more complicated plasmas, and the
resulting Bohm speed is consistent with the {\em underlying plasma
 transport model}. This last point accentuates the importance of an
accurate plasma transport model that properly accounts for the kinetic
nature of plasma transport within the Knudsen layer next to the wall,
not only for bulk plasma transport, but also for the Bohm sheath
constraint on wall-bound ion flow and energy flux.

We thank the U.S. Department of Energy Office of Fusion Energy
Sciences and Office of Advanced Scientific Computing Research for
support under the Tokamak Disruption Simulation (TDS) Scientific
Discovery through Advanced Computing (SciDAC) project at both
Virginia Tech under grant number DE-SC0018276 and Los Alamos National
Laboratory (LANL) under contract No.~89233218CNA000001. Yanzeng Zhang was supported under a Director's Postdoctoral Fellowship at LANL.
This research used resources of the National Energy Research
Scientific Computing Center (NERSC), a U.S. Department of Energy
Office of Science User Facility operated under Contract
No.DE-AC02-05CH11231. Useful discussions with Jun Li are acknowledged.

\nocite{*} 

\begin{thebibliography}{39}
\bibitem [{nom()}]{nominal-def-comment}%
  \BibitemOpen
  \href@noop {} {}\bibinfo {howpublished} {{The nominal sheath Knudsen number
  is computed with initial plasma parameters for an isotropic plasma before the
  sheath is dynamically formed with absorbing walls and an upstream source. The
  local sheath Knudsen number is computed with local plasma parameters after
  the steady-state sheath is established, so it varies with
  position.}}\BibitemShut {Stop}%
\bibitem [{clo()}]{closure-comment}%
  \BibitemOpen
  \href@noop {} {}\bibinfo {howpublished} {Chodura and Pohl's collisional
  closure captures the trend but is less accurate than closure flux from
  kinetic simulations in predicting the Bohm speed, as expected.}\BibitemShut
  {Stop}%
\end{thebibliography}
%

\end{document}



\title[Sample title]{Supplemental material to ``Bohm criterion of plasma sheaths away from asymptotic limits''}


\date{\today}

\maketitle

Here we provide supplemental information on the simulation setup,
presheath and sheath plasma profile, and the transport flux and Bohm
speed evaluated from first-principles particle-in-cell simulations.
Specifically the detailed profile for a slab plasma with a collisional
presheath from the kinetic simulation, shows a smooth transition
region between the presheath and the sheath, and demonstrates the
anisotropic nature of the plasma due to the decompressional cooling,
section~\ref{sec:setup-profile}. The temperature isotropization terms
$Q_{ee}$ and $Q_{ei}$ are calculated directly from the kinetic
simulation and compared with Chodura's closure, demonstrating the
needs for kinetic corrections to closure flux where local Knudsen
number is large, section~\ref{sec:T-relax}. Finally in
section~\ref{sec:Bohm-transition} we document the data comparison
between the predicted Bohm speed and the observed ion out flow speed
over a sheath transition region, and how different physical mechanisms
come together to set the Bohm speed.

\section{Simulation setup and steady-state plasma profile\label{sec:setup-profile}}

\begin{figure*}[!htp]
    \subfloat[]{\includegraphics[width=0.45\textwidth]{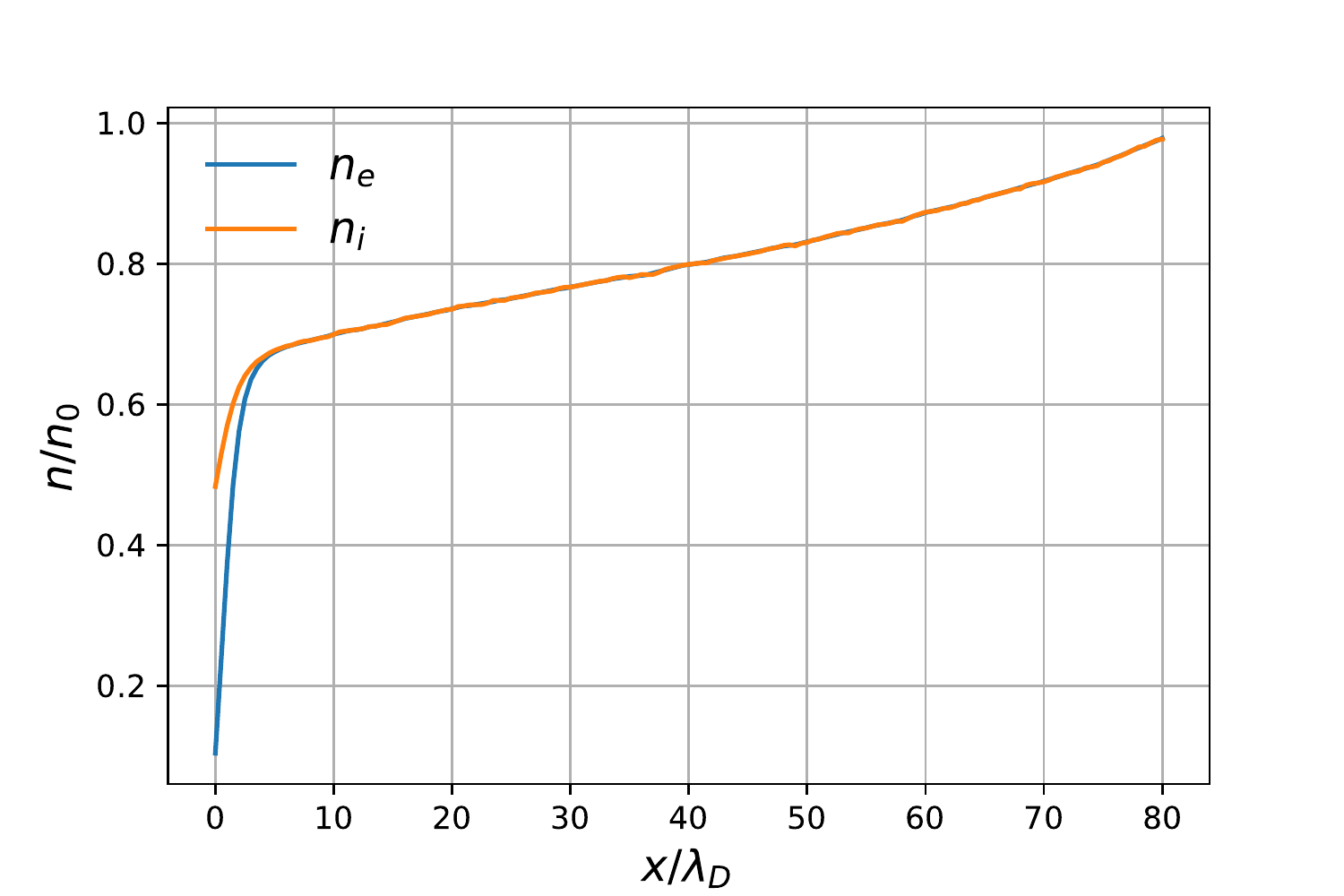}\label{fig:sheath-density}}
    \subfloat[]{\includegraphics[width=0.45\textwidth]{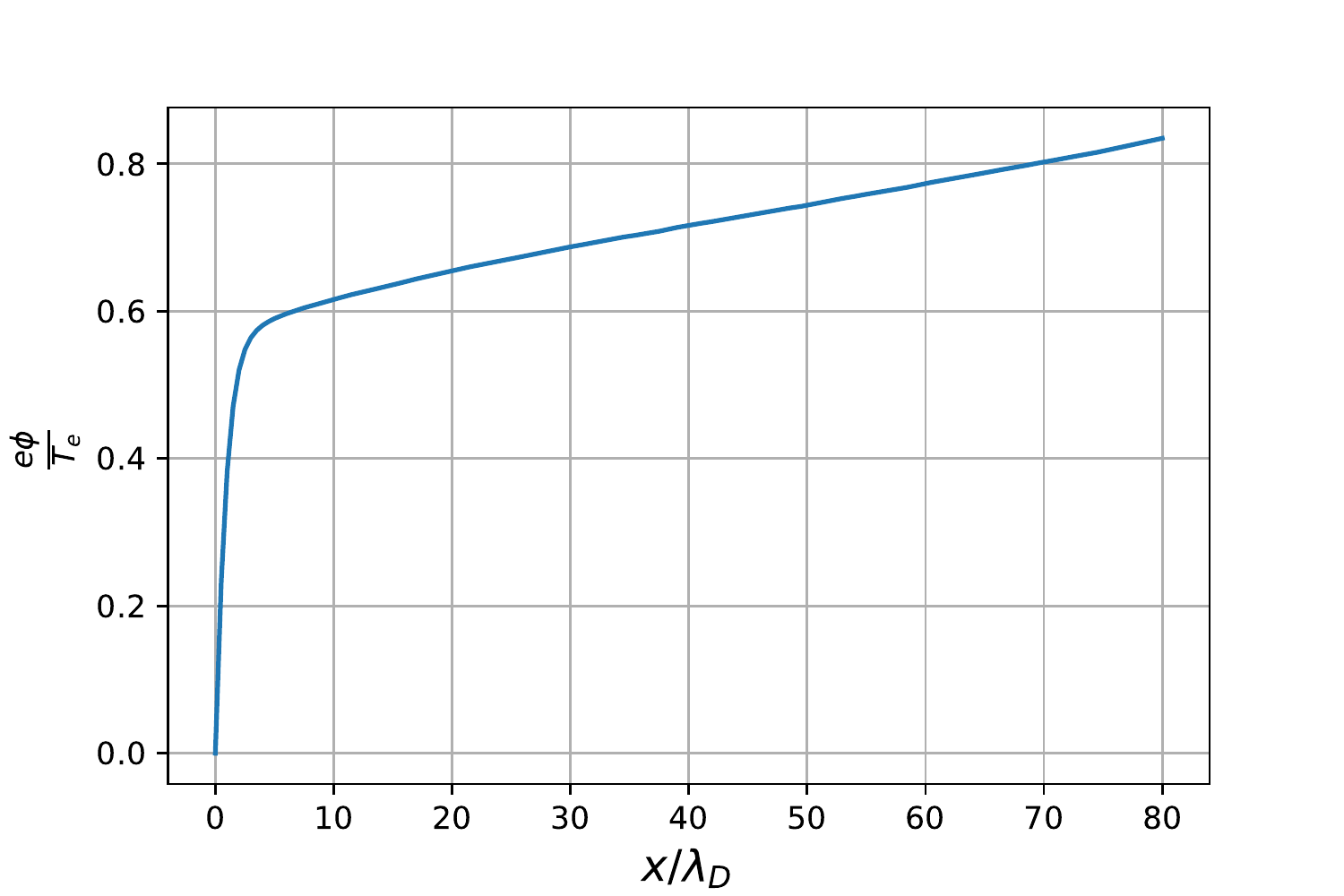}\label{fig:sheath-potential}}\\
    \subfloat[]{\includegraphics[width=0.45\textwidth]{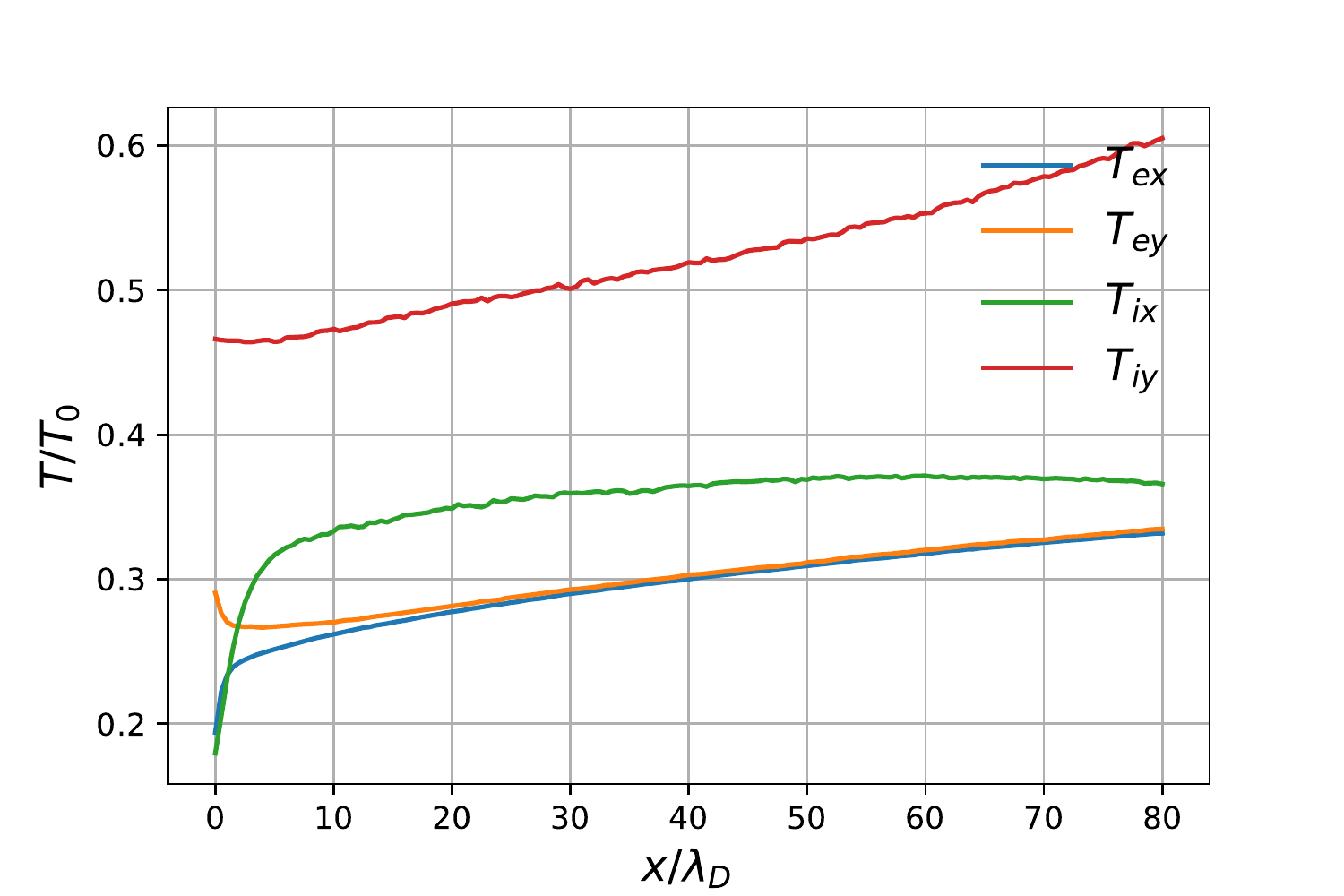}}
    \subfloat[]{\includegraphics[width=0.45\textwidth]{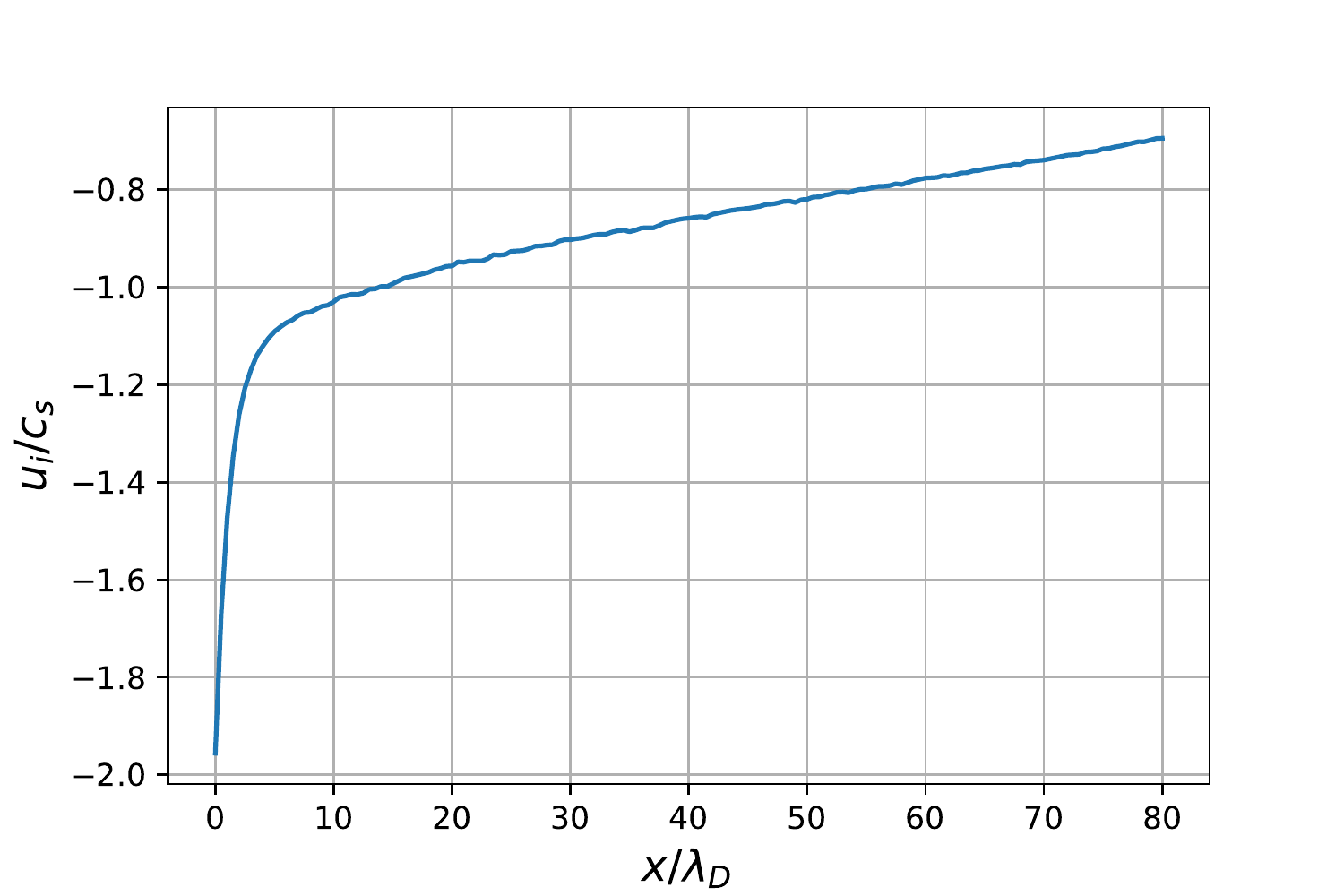}} 
    \caption{Steady-state presheath and sheath profile: (a) densities of
      electrons and ions normalized by the initial density $n_0$; (b)
      plasma potential normalized by $T_e/e$; (c) temperatures of
      electron and ion in $x$ (normal to the wall) and
      $y$(tangential to the wall) direction, normalized by the
      initial temperature $T_0$; (d) ion flow velocity normalized by the
      adiabatic sound speed $c_s = \sqrt{(T_{ex}+3T_{ix})/m_i}$;}
    \label{fig:sheath}
  \end{figure*}

Our simulations consider a slab plasma bounded by two absorbing walls, with
an upstream source in the middle that compensates the particle and
energy loss to the walls. The plasma parameters are averaged over
10000 time steps after reaching the steady state in order to reduce the
noise from the PIC simulation. The steady-state plasma profile for a
plasma with collisional presheath ($K_n \equiv \lambda_{mfp}/\lambda_D
= 75$) are shown in Fig.~\ref{fig:sheath}. The density and temperature
of both species are normalized to their initial values $n_0$ and
$T_0$, the potential $\phi$ to $T_0/e$, the ion flow to the local
adiabatic sound speed $c_s = \sqrt{(T_e+3T_i)/m_i}$.

 The electron temperature is anisotropic within the Knudsen layer (from
 the wall to $\lambda_{mfp}$, which is about $19.9\lambda_D$) due to
 decompressional cooling in the $x-$direction for a collisionless
 case\cite{Tang_2011}. Further upstream, the strong Coulomb collisions
 eliminate the electron temperature anisotropy. Strong ion temperature
 anisotropy extends much further upstream due to the slow ion-electron
 and ion-ion collision rate. In steady state, the plasma temperature
 gradient exists in the presheath and sheath region. The nominal sheath
 entrance is beyond $x = 5\lambda_D$, where the fractional
 charge gradient is $10\%$ and the ion exit flow is supersonic. These
 observations indicate that more involved transport physics need to be
 considered in the presheath-sheath transition problem.
  
\section{Comparison of temperature isotropization terms from kinetic simulation to Chodura's closure\label{sec:T-relax}}

The presheath/sheath plasma within a mean-free-path from the wall has
anisotropic temperature due to decompressional cooling in the flow
direction. Coulomb collisions among plasma particles can affect their
momentum and energy transport. The momentum transfer due to Coulomb
collisions between electrons and ions consists of two components, the
frictional force due to the net current $j$ and the thermal force
caused by the temperature gradient. Here we consider the cases with no
net current so frictional force can be neglected. In an anisotropic
plasma, the collisional energy exchange consists of temperature
equilibration between different species, relaxation of temperature
anisotropy, and Ohmic heating. In the presheath to sheath transition
problem, Ohmic heating is negligible in the absence of net current $j$
and the temperature equilibration between different species is
negligible due to the large ion-electron mass ratio. The dominant term
in energy exchange is temperature isotropization.  For the reason we
explained in the paper, despite not being a gradient term in the
plasma transport equation, it can play a critically important role in
Bohm criterion analysis, along with the more obvious leading order
terms in the multi-scale expansion of heat flux $q_n$ and thermal
force $R_T,$ which are gradient terms.

To verify that our newly derived Bohm speed agrees with direct
measurement of ion exit flow speed in the sheath transition region, we
compute heat flux, thermal force, and collisional isotropization terms
directly from kinetic simulation data.  Specifically, the momentum
energy transfer rate of species $a$ due to collisions with species $b$
can be evaluated from the summation over the particle population,
\begin{subequations}
  \begin{equation}
    \centering
    \label{eq:momt-change}
    \begin{aligned}
      I^{ab}_{\alpha} &= \int m_a v_\alpha\left(\ppx{f}{t}\right)^{ab}_{coll} d^3v \\
      & = \frac{N_f}{\Delta t}m_a \int \sum_{i=1}^{N_a}\left(\delta^3(v-v'_{i})-\delta^3(v-v_{i})\right) v_\alpha d^3v \\
      & = \frac{N_f}{\Delta t}m_a \sum_{i=1}^{N_a} \Delta v^{ab}_{i\alpha}
    \end{aligned}
  \end{equation}
  \begin{equation}
    \label{eq:energy-change}
    \begin{aligned}
      J_{\alpha\beta}^{ab} & = \int m_a (v_\alpha-V_\alpha)(v_\beta-V_\beta)\left(\ppx{f}{t}\right)^{ab}_{coll} d^3v \\
      &= \int m_a v_\alpha v_\beta\left(\ppx{f}{t}\right)^{ab}_{coll} d^3v -V_\alpha I_\alpha^{ab}-V_\beta I_\beta^{ab} + V_\alpha V_\beta S^{ab} \\
      &=  \frac{N_f}{\Delta t}m_a \sum_{i=1}^{N_a} (v^{\prime}_{i\alpha}v^{\prime}_{i\beta} - v_{i\alpha}v_{i\beta})  -V_\alpha I_\alpha^{ab}-V_\beta I_\beta^{ab}          
    \end{aligned}
  \end{equation}
\end{subequations}
where $\alpha$ and $\beta$ are directions, $N_f$ is a normalizing
factor and $N_a$ is the total number of super particles of species $a$
in the simulation. $S^{ab}$ is the density change rate and it equals
$0$ for Coulomb collisions. {Here $(\partial
  f/\partial t)_{coll}^{ab}$ is the change of the distribution
  function due to the Coulomb collision between species $a$ and $b$
  and it is evaluated using Takizuka and Abe's method~\cite{T.A} in
  VPIC. It can be noted that Takizuka and Abe's method ensures
  momentum and energy conservation in binary collisions, to the
  machine accuracy in floating point calculations on a computer.}  In the case of no
net current and slow temperature relaxation rate between different
species, thermal force $R_T$ and the temperature isotropization $Q$
can be directly obtained from Eq.~\ref{eq:momt-change} and
Eq.~\ref{eq:energy-change} in a PIC simulation.

\begin{figure*}[!htp]
  \subfloat[]{\includegraphics[width=0.45\textwidth]{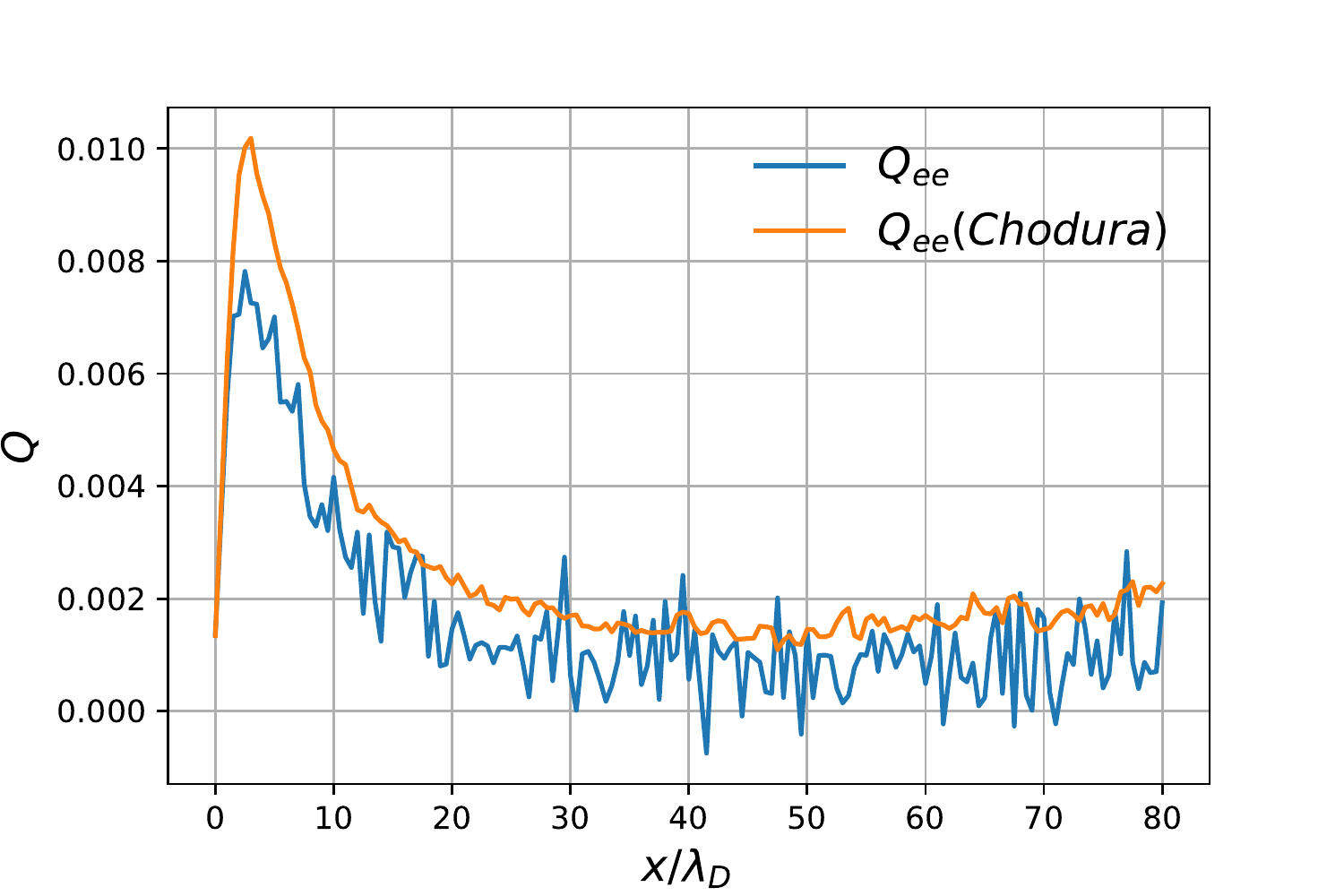}\label{fig:density}}
  \subfloat[]{\includegraphics[width=0.45\textwidth]{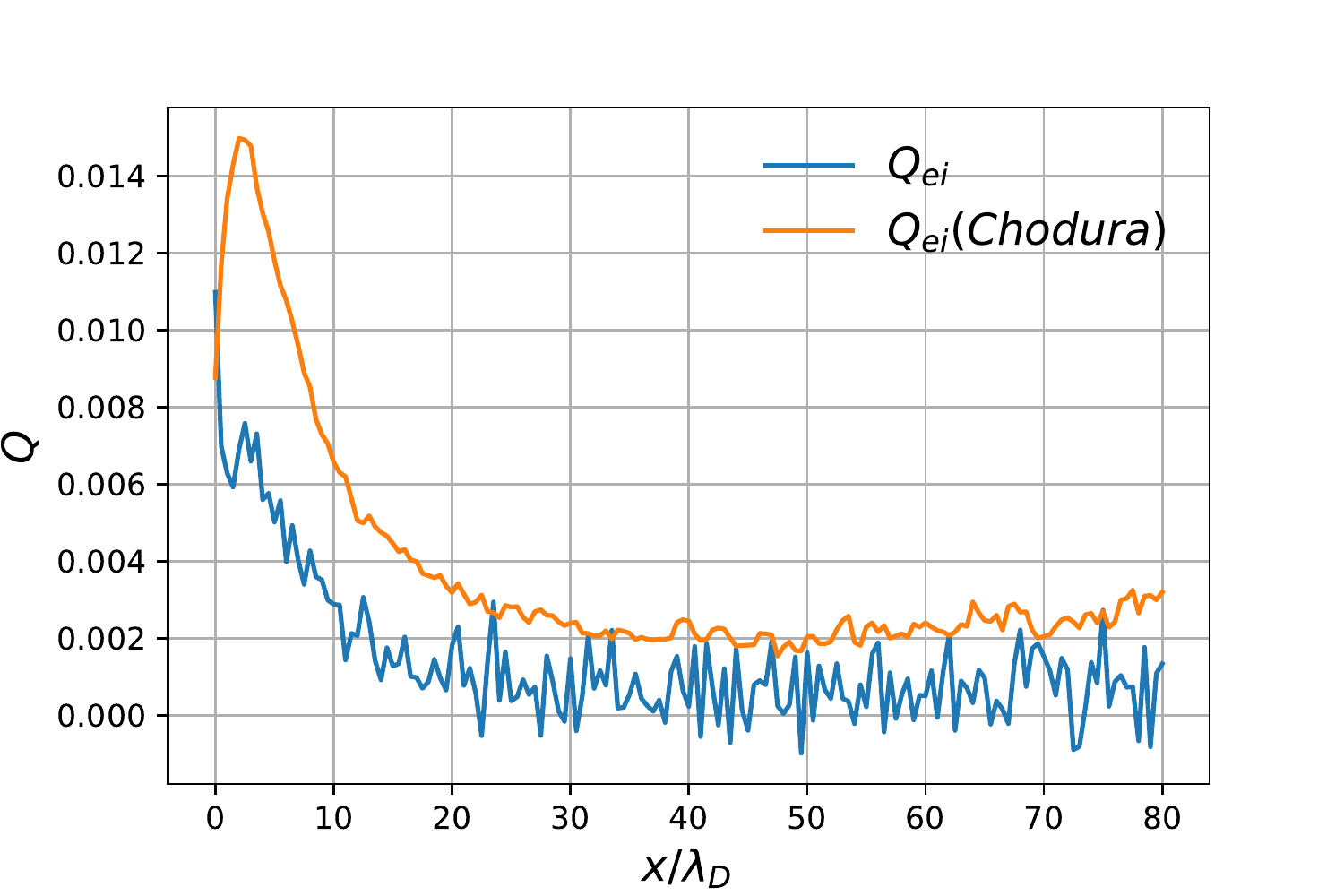}\label{fig:potential}}\\
  \caption{Comparison of $Q_{ee}$ and $Q_{ei}$ terms evaluated from kinetic simulation with those calculated by Chodura's closure for a collisional presheath with $K_n = 75$.}
  \label{fig:compare-Q} 
\end{figure*}

In the collisional limit, Chodura and Pohl provide a closure for the
anisotropic plasma transport~\cite{Chodura_1971}. In present analysis,
we focus on cases where the collisional mean-free-path is much longer
than the Debye length. Furthermore, the plasma distribution functions
are skewed by the free-streaming loss so kinetic corrections to
Chodura's closure are expected. Here, we contrast the temperature
isotropization terms $Q_{ee, ei}$ obtained by
Eq.~\ref{eq:energy-change} with those calculated from Chodura's
expression (see Eq.$~6$ in the paper) in Fig.~\ref{fig:compare-Q}

From Fig.~\ref{fig:compare-Q}, one can see that while Chodura's
closure captures the underlying collisional isotropization physics
admirably, kinetic correction in $Q_{ei}$ is important for accurate
evaluation of Bohm speed using Eq.~(12) of the paper.

\section{Bohm speed variation in the sheath transition region~\label{sec:Bohm-transition}}

To show the validity of the Bohm speed model over a sheath transition
region rather than one single point at the sheath entrance, aside from
Fig.~(2) in the paper, here we provide the value of the physical
quantities for evaluating the Bohm speed in Eq.~(12) of the paper over
a small region $x\in[5.0\lambda_D, 10.0\lambda_D]$ near the sheath
entrance in Table.\ref{tab:normalized-parameters}.

\begin{table}[H]
\caption{\label{tab:normalized-parameters} Normalized parameters determining the Bohm speed in the sheath transition region for $K_n = 200$. }
\begin{tabular}{ccccccccccccccc}
  $\cfrac{x}{\lambda_D}$ & $\cfrac{u_{ix}^{se}}{c_s}$ & $\cfrac{u_{Bohm}}{c_s}$ & $ \cfrac{1}{e\Gamma_e^{se}} \cfrac{\partial q_n^{e}}{\partial\phi}$ & $\cfrac{Q_{ee}^{se}+Q_{ei}^{se}}{e\Gamma_{e}^{se}E^{se}}$ & $ \cfrac{1}{e\Gamma_e^{se}} \cfrac{\partial q_n^{i}}{\partial\phi} $& $\cfrac{Q_{ii}^{se}}{e\Gamma_{e}^{se}E^{se}} $& $\alpha$\\
  \hline
  5.0 & 1.04 & 1.01 & -1.67 & 3.05 & -0.21 & 0.33 & 0.39 \\
  5.5 & 1.03 & 1.00 & -2.20 & 3.55 & -0.22 & 0.38 & 0.45 \\
  6.0 & 1.02 & 0.99 & -2.77 & 4.02 & -0.24 & 0.46 & 0.49 \\
  6.5 & 1.01 & 1.00 & -3.15 & 4.11 & -0.24 & 0.48 & 0.44 \\
  7.0 & 1.00 & 0.99 & -3.23 & 4.33 & -0.27 & 0.55 & 0.47 \\
  7.5 & 0.99 & 1.00 & -3.46 & 4.57 & -0.30 & 0.54 & 0.49 \\
  8.0 & 0.99 & 0.99 & -3.92 & 4.72 & -0.20 & 0.54 & 0.41 \\
  8.5 & 0.98 & 0.97 & -4.19 & 5.03 & -0.17 & 0.63 & 0.44 \\
  9.0 & 0.98 & 0.96 & -4.38 & 5.35 & -0.20 & 0.66 & 0.51 \\
  9.5 & 0.97 & 0.97 & -4.45 & 5.40 & -0.22 & 0.63 & 0.49 \\
  10.0& 0.97 & 0.97 & -4.11 & 4.93 & -0.25 & 0.70 & 0.50 \\
\end{tabular}
\end{table}

The model gives an accurate prediction for the Bohm speed over the
sheath transition region as is shown by comparing the second and third
column in Table.\ref{tab:normalized-parameters}. Even though the
normalized physical quantities, such as heat flux and temperature
isotropization rate, vary enormously in the sheath transition region,
those large changes compensate with each other to capture the slowly
varying ion exit flow speed or Bohm speed in the sheath transition
region.
  
%